\documentclass[12pt,a4paper,final]{iopart}

\usepackage{iopams}  
\usepackage{graphicx}
\usepackage{cite}
\usepackage{hyperref}
\hypersetup{breaklinks=true,colorlinks=true,linkcolor=blue,urlcolor=blue,citecolor=blue}

\begin{document}

\title[Influence of surface materials on the volume production of negative ions]{Influence of surface materials on the volume production of negative ions in a radio-frequency driven hydrogen plasma}

\author{J. Ellis$^{1,2}$}
\address{$^1$York Plasma Institute, Department of Physics, University of York, York, YO10 5DD, U.K. \newline Now at: $^2$Leibniz Institute for Plasma Science and Technology (INP), Felix-Hausdorff-Str. 2, 17489 Greifswald, Germany}
\ead{james.ellis@inp-greifswald.de}

\author{J. Branson, K. Niemi, E. Wagenaars, T. Gans}
\address{York Plasma Institute, Department of Physics, University of York, York, YO10 5DD, U.K.}

\begin{abstract}
Negative atomic hydrogen ion (H$^{-}$) densities were measured in a pulsed low-pressure E-mode inductively-coupled radio-frequency (rf) driven plasma in hydrogen by means of laser photodetachment and a Langmuir probe. This investigation focuses on the influence of different metallic surface materials on the volume production of H$^{-}$ ions. The H$^{-}$ density was measured above a thin disc of either tungsten, stainless steel, copper, aluminium, or molybdenum placed onto the lower grounded electrode of the plasma device as a function of gas pressure and applied rf power. For copper, aluminium, and molybdenum the H$^{-}$ density was found to be quite insensitive to pressure and rf power, with values ranging between 3.6x10$^{14}$ to 5.8x10$^{14}$ m$^{-3}$. For stainless steel and tungsten, the H$^{-}$ dependency was found to be complex, apart from the case of a similar linear increase from 2.9x10$^{14}$ to 1.1x10$^{15}$ m$^{-3}$ with rf power at a pressure of 25 Pa. Two-photon absorption laser induced fluorescence was used to measure the atomic hydrogen densities and phase resolved optical emission spectroscopy was used to investigate whether the plasma dynamics were surface dependent. An explanation for the observed differences between the two sets of investigated materials is given in terms of surface reaction mechanisms for the creation of vibrationally excited hydrogen molecules.
\end{abstract}

\section{Introduction}

The study of negative ions, in low pressure plasmas, has become increasingly important. In particular, for applications such as: plasma etching \cite{Shindo_1995}, plasma thrusters for propulsion \cite{6990637}, and neutral beam injection \cite{KURIYAMA1995445, doi:10.1063/1.2814248, doi:10.1063/1.3541790}. Additionally, negative ions are important to global plasma dynamics \cite{Babkina_2005, Schiesko_2008} and therefore should be investigated in order to ascertain their influence on the plasma bulk. A key negative ion species, for neutral beam injection, is H$^{-}$ (or D$^{-}$) due to its importance in magnetically confined fusion reactors. 

Negative hydrogen ions are generated through two key mechanisms. Firstly, volume production of negative ions \cite{doi:10.1063/1.4921298} which relies on the dissociative attachment of vibrationally excited molecules with impinging electrons:

\begin{equation}
H_{2}(v) + e^{-} \rightarrow H^{-} + H
\end{equation}

Secondly, surface production of negative ions \cite{FANTZ20127}. This process results from the capture of an electron upon an impinging atomic species at a surface, or, the capture of two electrons by an impinging singularly charged positive ion. Each of these mechanisms have distinct advantages and disadvantages, depending on the desired application of the negative ions. For instance, surface production is preferred for the production of a large quantity of negative ions for neutral beam injection; this is because the ions must subsequently be extracted from the plasma \cite{FANTZ20127, Franzen_2011}.

There are a number of experimental studies on the effect that the surface has on the direct production of H$^{-}$ ions \cite{FRANZEN20133132, doi:10.1063/1.4932560, doi:10.1063/1.5012591}. For example, the ITER negative ion source, developed at IPP Garching, utilises the deposition of caesium onto a metallic grid in order to lower the materials' work function. This causes a highly efficient production of negative ions at the surface from the capture of liberated electrons. However, there are many drawbacks to the use of caesium within a negative ion source, for instance, regular cleaning is necessary in order to maintain a consistent beam profile. Other materials are being investigated for negative ion production, in an attempt to mitigate these problems, including highly oriented pyrolytic graphite and diamond \cite{Cartry_2017, Sasao_2018, doi:10.1063/1.3258352}. 

The importance of the surface on both plasma dynamics and chemistry is well established \cite{doi:10.1063/1.4841675, Greb_2015, doi:10.1063/1.4979855, Osiac_2007}. Direct surface production of negative ions is only one example of a plasma surface interaction that may influence the negative ion density, there are other mechanisms which can have indirect consequences.  For example, atomic species, which can be lost through various interactions with a wall.  Firstly, when an atom impinges a surface, it will gain energy due to the interaction of the unbound electron and the surface which forms a chemical bond. Typically, the energy gained is approximately 2.4 eV for hydrogen and metallic surfaces \cite{CHRISTMANN19881}.  These adsorbed atoms can then undergo different mechanisms for their release, as depicted by Figure 1. The Langmuir-Hinshelwood (LH) \cite{CHRISTMANN19881} recombination process involves two adsorbed atoms recombining to create a molecule. The H-H bond energy is approximately 4.5 eV, making this process endothermic by 0.3 eV. The second mechanism, known as Eley-Rideal (ER) recombination \cite{doi:10.1063/1.467776, doi:10.1063/1.474302}, involves an impinging atom reacting directly with a previously adsorbed atom, again creating a molecular species. This process is exothermic by approximately 2 eV, as only one atom is adsorbed onto the surface during the recombination process. These two mechanisms are idealised, as an impinging atom requires time to reach thermal equilibrium with the surface. There is a third surface recombination process, in which the impinging atom is first weakly bound to the surface and is able to move relatively easily across the surface; this allows energy to be transferred to other absorbed species. This process was first proposed by Harris and Kasemo as 'Hot Atom' (HA) recombination \cite{HARRIS1981L281}. Since this paper, multiple studies have given evidence that the ER mechanism alone cannot account for observations \cite{doi:10.1063/1.480145, KAMMLER200091}; providing indirect support for the relevance of the HA process.

\begin{figure}[htb!]
 \centering
 \includegraphics[width=12cm]{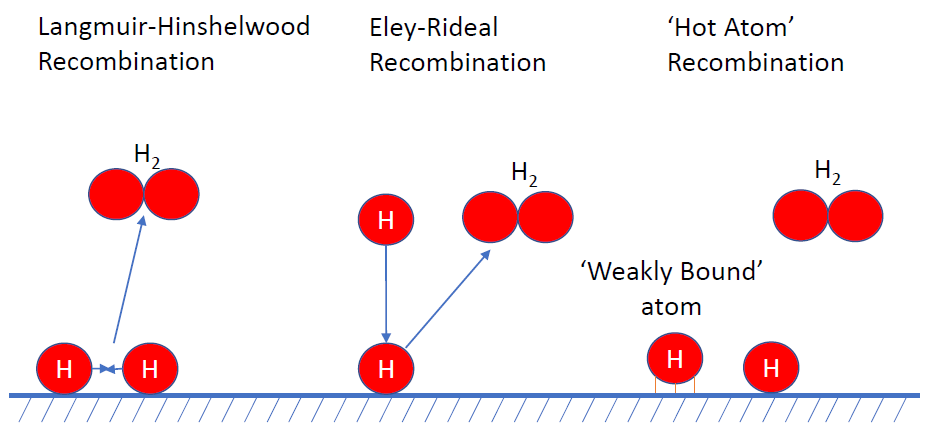}
 \caption{\label{figureone}Depicted here are three types of atom-surface interactions. From left to right: Langmuir-Hinshelwood recombination; two adsorbed atomic hydrogen particles can recombine producing a molecular hydrogen, Eley-Rideal recombination; an impinging atomic hydrogen directly reacting with an adsorbed hydrogen atom producing molecular hydrogen, and 'Hot Atom' recombination; a weakly bound hydrogen atom reacts with an adsorbed atom producing molecular hydrogen.}
\end{figure}

The two exothermic recombination mechanisms are ER and HA, with the excess energy being distributed into various degrees of freedom, including vibrational energy levels. The production of vibrationally excited molecular hydrogen has been found to be strongly surface dependent \cite{PhysRevLett.60.337, doi:10.1063/1.468242}. Whilst these mechanisms are not well understood, a study by Markelj \cite{doi:10.1063/1.3569562} proposes that they may be driven through weak binding sites on the surface. 

This paper investigates the production of H$^{-}$ ions in a low pressure inductively coupled hydrogen plasma source operated in the electrostatic mode (E-mode). Multiple metallic surfaces were installed on the grounded electrode, and the H$^{-}$ densities were measured by use of laser photodetachment. This was conducted in order to ascertain the influence of metallic surfaces upon the volume production of H$^{-}$ ions. Furthermore, atomic hydrogen densities were measured by use of two-photon absorption laser induced fluorescence (TALIF) to investigate the possible influence of atomic hydrogen on the H$^{-}$ ion densities. Finally, phase resolved optical emission spectroscopy (PROES) was used in order to determine whether the surface significantly altered the plasma dynamics.

\section{Experimental Description}

The source chamber used was a modified Gaseous Electronics Conference (GEC) reference cell \cite{doi:10.1063/1.1144770}, operated as an E-mode inductively coupled plasma. A driving frequency of 13.56 MHz was supplied by the power generator (Advanced Energy, Cesar 1310) into an automatic matching network (Advanced Energy, RF Navio) affixed to the top of the cell with a direct connection to a 5-turn copper coil. In combination with a digital delay generator (Stanford Research Systems, DG 645), a pulsed plasma was generated at a repetition rate of 10 Hz with a 10~\% duty cycle. The plasma was pulsed to ensure a consistent temperature between measurements, because it is known that surface processes can be temperature dependent \cite{Abdel_Rahman_2005}. A feed gas of hydrogen (99.9995~\% purity) was flowed through a mass flow controller and the pressure was controlled using a butterfly valve to achieve pressures in the range of 17--37.5 Pa. The plasma was operated exclusively in the E-mode regime, while the inductively coupled H-mode regime at applied powers above 800 W is not covered in the present investigation; as the Langmuir probe was unable to withstand the higher temperatures in the H-mode regime.

The electrode spacing was measured to be 42 mm between the top quartz window and the bottom grounded electrode; the walls of the chambers were also grounded. Different material samples were installed onto the bottom grounded electrode: tungsten, copper, aluminium, molybdenum, and stainless steel 304. These samples were of uniform thickness (3.1 mm), with a diameter of 104 mm, covering the diameter of the bottom electrode without touching the surrounding guard ring. The pulsed operation of the plasma was chosen to reduce the thermal heating of the installed sample. 

In order to measure the negative ion densities, laser photodetachment was conducted. The used Q-switched Nd:YAG laser with frequency doubling option (Continuum, Minilite II) provided narrow-band radiation at 532 nm wavelength in pulses of 20 mJ energy within a 5 ns duration at a 10 Hz repetition rate. The laser output beam of 3 mm diameter was aligned along the dog-legged part of the Langmuir probe tip (tungsten wire of 0.15 mm diameter along 9 mm). The probe tip was located 9 mm above the central point of the installed sample, approximately 30 mm below the powered electrode. The probe tip was connected into an insulating alumina shaft, which allowed the signal to be routed through filtering components located away from the probe tip.  

In the experimental procedure the Langmuir probe was first used to acquire an IV-curve through the use of a dedicated Hiden ESPion controller. From this IV-curve, the plasma potential (V$_p$) was calculated by using the crossing point of the second differential. For the photodetachment measurements, the probe was biased above the plasma potential, in order to ensure that the majority of photodetached electrons from the laser interaction volume were collected. This additional bias was chosen to be 7 V which proved large enough to be within the electron saturation regime, whilst not creating measurable secondary emission effects. The current corresponding to the applied voltage (V$_{app}$) was then recorded from the IV-curve. In order to measure the photodetachment signal, the Hiden ESPion controller was then disconnected and the probe connected to a DC power supply (HP 6115A 0-50 V) and the DC 1 M$\Omega$ output of a Teledyne Lecroy Wavesurfer 3054 oscilloscope. The analysis circuitry for this experiment was based on J. Santoso et. al. \cite{doi:10.1063/1.4931469}. The laser pulse was synchronised to occur 2 ms after the plasma ignition, after which the plasma was observed to have reached a quasi-steady state. The probe was cleaned regularly to avoid adverse effects of deposits on the tip. The cleaning procedure involved positively biasing (60 V) the tip in a 500 W continuously driven hydrogen plasma.

Once the photodetachment pulse was measured, it was averaged over 128 scans to reduce background noise. The maxima of the photodetachment pulse can be compared to the current at V$_{app}$ to calculate the H$^{-}$ fraction, with respect to the electron density ($\frac{\Delta I}{I}$). In order to find absolute densities, the electron density must be known. The latter quantity is calculated using the Hiden ESPsoft software package based on orbital motion limited probe theory \cite{PhysRev.28.727, Allen_1992}. Whilst this is possible, it was typically less reliable than finding the positive ion density (n$_{i}$) from the ion saturation regime. Which, under the assumption of quasi-neutrality, can be used to find the absolute negative ion density as shown by equation 2. It is worth noting that RF-driven plasmas of this pressure tend to be dominated by H$_{3}^{+}$ ions \cite{Samuell_2015}, therefore the mass number three was used during the analysis routine. 

\begin{equation}
n_{H^-} = \frac{\Delta I}{I}.\frac{n_i}{(1 + \frac{\Delta I}{I})}
\end{equation}

In addition, PROES was applied to check whether secondary electron emission from the sample surface can play a major role for the H$^{-}$ formation under the investigated conditions. The optical plasma emission, as a function of the phase within the rf excitation cycle, was detected with an iCCD camera (Andor iStar DH344T-18U-73), with a 656 nm interference filter installed (to provide sensitivity for the hydrogen $\alpha$ Balmer line), and a digital delay generator. This emission is then combined with the effective lifetime of the upper level in order to calculate the excitation rate. In this instance, the effective lifetimes were measured, using the same apparatus required for TALIF which is discussed next, to be approximately 12 ns. For more information regarding the diagnostic technique please refer to Gans et. al. \cite{PhysRevA.67.012707}.

TALIF was used to measure the effective lifetimes, as mentioned previously, and the atomic hydrogen densities. The theory of TALIF is covered in greater detail elsewhere \cite{doi:10.1063/1.476388, doi:10.1063/1.1425777}, however, a brief description will be given here. The experimental set up comprised of a 10 Hz, ns-pulsed (3-5 ns) laser (Continuum Surelite, SL-EX Nd:YAG, coupled with a Continuum Horizon optical parametric oscillator) capable of lasing between 192--2750 nm. 

The energy of the laser was attenuated before entering the plasma; this is done to ensure the energy is low enough to cause negligible photoionisation events. This was confirmed by measuring the saturation of the fluorescence signal as a function of the square of the laser energy. The laser energy which was used for all the TALIF measurements was 780 $\mu$J for hydrogen, and 180 $\mu$J for krypton. The focal point of the laser was in the centre of the chamber, approximately 1 cm above the grounded electrode. The fluorescence signal was captured by the aforementioned iCCD camera, with an appropriate interference filter installed to allow sensitivity to the fluorescence wavelength (656 nm for hydrogen and 826 nm for krypton). A temporally (200 ns) and spectrally integrated (0.01 nm) measurement was conducted to measure the total fluorescence signal in each instance. This was combined with a temporally resolved measurement, that allowed the effective decay rate to be determined.

Measuring the total fluorescence signal as a function of wavelength and time only allows for qualitative measurements, unless specific information about the lasers spatio-temporal characteristics are known. An alternative method for calculating absolute densities requires a calibration of the total fluorescence signal. This can be conducted in a number of ways, however, in this work the calibration was performed using krypton as described by Niemi et. al. \cite{Niemi_2001}. Figure 2 shows the excitation scheme, and fluorescence, for atomic hydrogen with respect to krypton.

\begin{figure}[htb!]
 \centering
 \includegraphics[width=6cm]{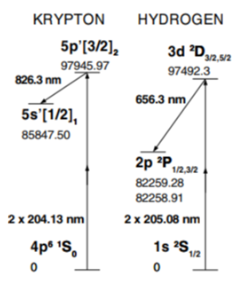}
 \caption{\label{figureeight}TALIF excitation, and fluorescence, scheme for atomic hydrogen with respect to krypton \cite{Niemi_2001}.}
\end{figure} 

\section{Experimental Results}

Figure 3 shows the measured H$^{-}$ density as a function of the applied rf power at a fixed pressure of 25 Pa for all five investigated surface materials. The five surfaces are clearly separated into two distinct groups, one containing steel and tungsten, the other containing copper, aluminium, and molybdenum. The steel and tungsten samples show a linear increase in the H$^{-}$ density with power, whereas, the copper, aluminium, and molybdenum samples are mostly unchanged with increasing power. 

\begin{figure}[htb!]
 \centering
 \includegraphics[width=12cm]{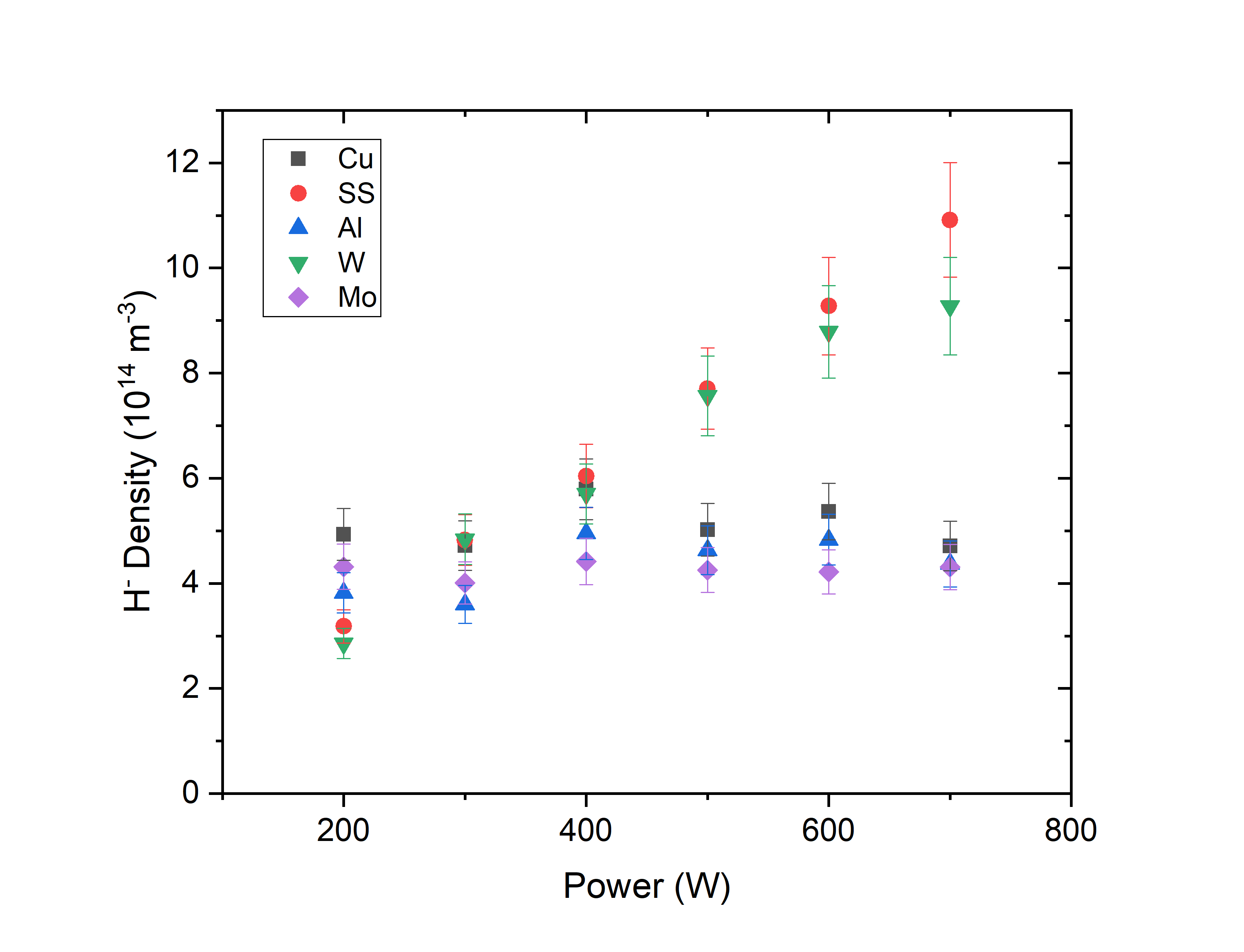}
 \caption{\label{figureeight}Measured H$^{-}$ density as a function of applied rf power for different surface materials at a fixed pressure of 25 Pa.}
\end{figure} 

There are a number of ways in which the surface can affect the bulk plasma including: secondary electron emission, direct surface production of negative ions, and recombination of atomic species. Each of these mechanisms may affect the measured negative ion density and therefore should be discussed. 

Secondary electron emission creates electrons at the surface that are subsequently accelerated into the bulk plasma, these energetic electrons (hereafter named $\gamma$ electrons) may create vibrationally excited species of molecular hydrogen. The dissociative attachment process, as outlined by Equation 1, requires vibrationally excited molecular species in order to generate negative ions. If the number of secondary electrons varied between the steel and copper surfaces, then this could explain the observed difference in the H$^{-}$ densities. Figure 4 compares the PROES measurements between a copper (left) and steel (right) surface. The driven electrode is located at the top of the images (4 cm), and the grounded metallic surface is located at the bottom (0.8 cm). The plasma is operated at the same rf power of 500 W and pressure of 20 Pa, chosen as the difference in the H$^{-}$ densities were larger for 20 Pa (shown later) than 25 Pa, for both of these images. The PROES images imply that the plasma dynamics are almost identical in each case. If the difference in H$^{-}$ densities were due to $\gamma$ electrons, this should be observed as a differing intensity between the surfaces. The instance of this peak should be when the sheath is fully expanded (\textbf{II}), however, this is not observed. Examples of $\gamma$ electron peaks can be found in the following references \cite{Doyle_2018, Abdel_Rahman_2007}.

\begin{figure}[htb!]
 \centering
 \includegraphics[width=14cm]{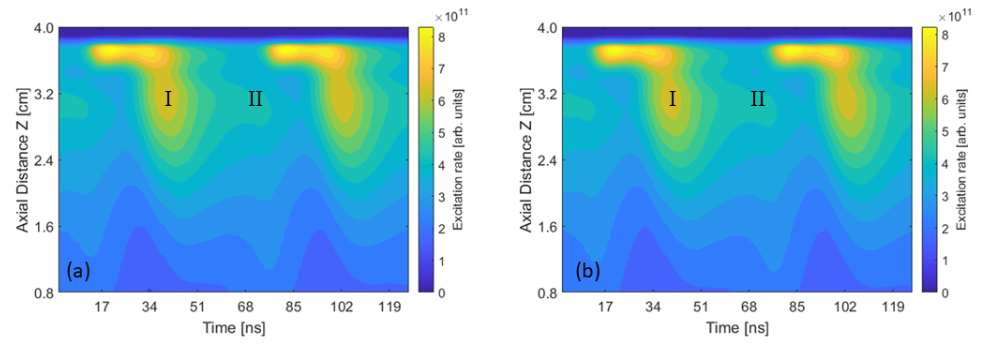}
 \caption{\label{figurefive}Results of phase resolved optical emission spectroscopy measurements showing the excitation rate of the hydrogen $\alpha$ Balmer line at 656 nm. The plasma is operated at 20 Pa with an applied power of 500 W. Image (a) is for the copper sample installed, and the image (b) is for the stainless steel 304 sample. \textbf{I} denotes the instance of sheath expansion, and \textbf{II} denotes when the sheath is fully expanded. During \textbf{II} the effect of $\gamma$ electrons should be the most pronounced. }
\end{figure}

Direct surface production of negative ions largely depends on the work function of the material \cite{Cartry_2017}, because a lower work function allows an electron to be more easily liberated from the surface. This electron can in turn be captured by an atomic species, or, two electrons can be simultaneously captured by a singularly charged positive ion. The work functions of the samples are: 4.55, 4.3, 4.65, 4.28, and 4.6 eV for tungsten, stainless steel 304, copper, aluminium, and molybdenum, respectively \cite{doi:10.1063/1.323539, doi:10.1063/1.1708797}. The similarities between these work functions implies that direct surface production of negative ions is unlikely the cause of these discrepancies in H$^{-}$ density. 

The final consideration to make is the recombination of atomic species at the surface. As previously discussed, there are a number of pathways from which atomic recombination at the surface can occur. Both the ER, and HA recombination process can produce vibrationally excited molecular species due to their exothermic nature. These vibrationally excited species may then be released into the plasma where dissociative attachment can take place, producing negative ions.

These surface produced vibrationally excited species may provide the answer for surface to surface variation in the measured H$^{-}$ densities. Whilst we were unable to find direct studies comparing the vibrational distribution of all the surfaces studied, there is a study by S. Markelj and I. Cadez, in which they compared the vibrational distribution of molecular hydrogen between a copper and tungsten surface \cite{doi:10.1063/1.3569562}. They observed a much larger vibrational temperature with a tungsten surface as opposed to a copper surface. They combined their experimental measurements with a Monte Carlo simulation which suggested the difference in vibrational temperature may be due to the presence of weak binding sites on the surface of tungsten \cite{doi:10.1063/1.1674753}. As the cross-section for dissociative attachment increases with increasing vibrational energy level \cite{PhysRevA.29.106} this is the most likely explanation why the measured H$^{-}$ density is larger for a tungsten surface installed as opposed to a copper surface. It has also been shown separately that stainless steel increases the relative density of H$^{-}$ by a factor of two over a copper surface \cite{doi:10.1063/1.448668}.

\begin{figure}[htb!]
 \centering
 \includegraphics[width=12cm]{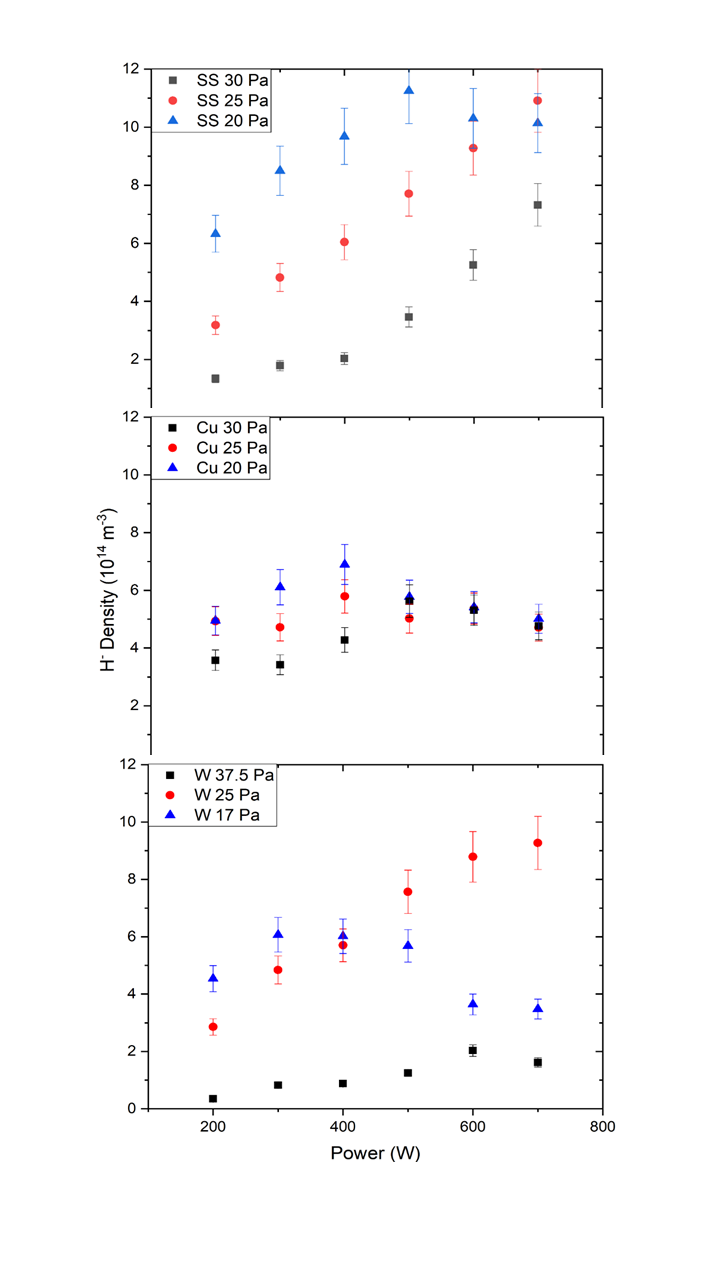}
 \caption{\label{figureseven}Measured H$^{-}$ density as a function of applied rf power for the steel, copper, and tungsten surfaces installed at three different pressures.}
\end{figure}

Figure 5 shows the H$^{-}$ density as a function of power for the steel, copper, and tungsten samples installed. For all three samples, increasing the pressure results in a lower initial H$^{-}$ density. The tungsten measurements were conducted at a higher (37.5 Pa), and lower pressure (17 Pa) than that of steel. The 30 Pa measurement with the steel sample shows an unchanged H$^{-}$ density between 200 and 400 W. A similar result can be observed for the tungsten sample at 37.5 Pa. The lower pressure results, 20 Pa and 17 Pa, for steel and tungsten respectively, show an initially high H$^{-}$ density which increases with power until a maxima. A subsequent increase in power past this maxima results in a decrease in the H$^{-}$ density. The 17 Pa measurement for tungsten shows this maxima occurring at lower power than for the steel 20 Pa measurement, which may be attributed to the destruction of negative ions through electron detachment. The copper sample shows a similar result, with the highest pressure (30 Pa) resulting in the lowest H$^{-}$ density and the lowest pressure (20 Pa) yielding the highest density. However, the range of the densities is much smaller between these pressures, and the H$^{-}$ densities converge towards powers above 500 W. This implies that the surface produced vibrationally excited molecules are not as prominent with a copper surface installed. 

It has been observed that the differences in the H$^{-}$ densities between pressures depends on the surface, this may be due to a number of mechanisms. Firstly, we must consider the fast electrons which are required in order to create vibrationally excited species, a necessary precursor for volume produced negative ions. Secondly, we can consider the associative detachment pathway, through which a H$^{-}$ ion can be destroyed via a collision with an atomic hydrogen. However, if these were the only mechanisms then the variation of the H$^{-}$ density with pressure should be similar between the surfaces. Interestingly, that is not what has been observed; the variation in H$^{-}$ densities is more pronounced with the steel and tungsten sample installed than with the copper sample. One possible explanation may be that of vibrational relaxation due to collisions with atomic hydrogen. If a surface produces more vibrationally excited species than another, then the effect of vibrational relaxation on the H$^{-}$ density will be more pronounced. 

Given the importance of the atomic hydrogen density on the creation and destruction of H$^{-}$ ions, it was prudent to investigate how this quantity scaled with increasing pressure. Figure 6 shows the atomic hydrogen densities, as measured by TALIF, as a function of power for three different pressures, with the copper sample installed. Increasing the gas pressure for a set power results in an increase in the atomic hydrogen densities; the reason for this is not necessarily trivial. Firstly, as the pressure increases, for a constant volume and temperature, the number of molecules will increase. Equally, the electron density increases with pressure, this means that more more electron-molecule collisions can occur, which may result in dissociation. Secondly, as the pressure is increased the number of collisions rises; this acts to reduce the electron temperature, which in turn reduces the number of electrons with sufficient energy to cause dissociation. Finally, the primary loss mechanism for atomic hydrogen is due to the diffusion to the walls, and it has been previously reported that increasing the pressure acts to decrease the surface losses \cite{doi:10.1063/1.1490630}. Due to the complexity of all these factors, it was necessary to measure the atomic hydrogen density in order to ascertain the dependency on pressure. The absolute atomic hydrogen densities obtained within this investigation, and qualitative trends, are comparable to previous studies under analogous conditions \cite{Abdel_Rahman_2006}.

\begin{figure}[htb!]
 \centering
 \includegraphics[width=12cm]{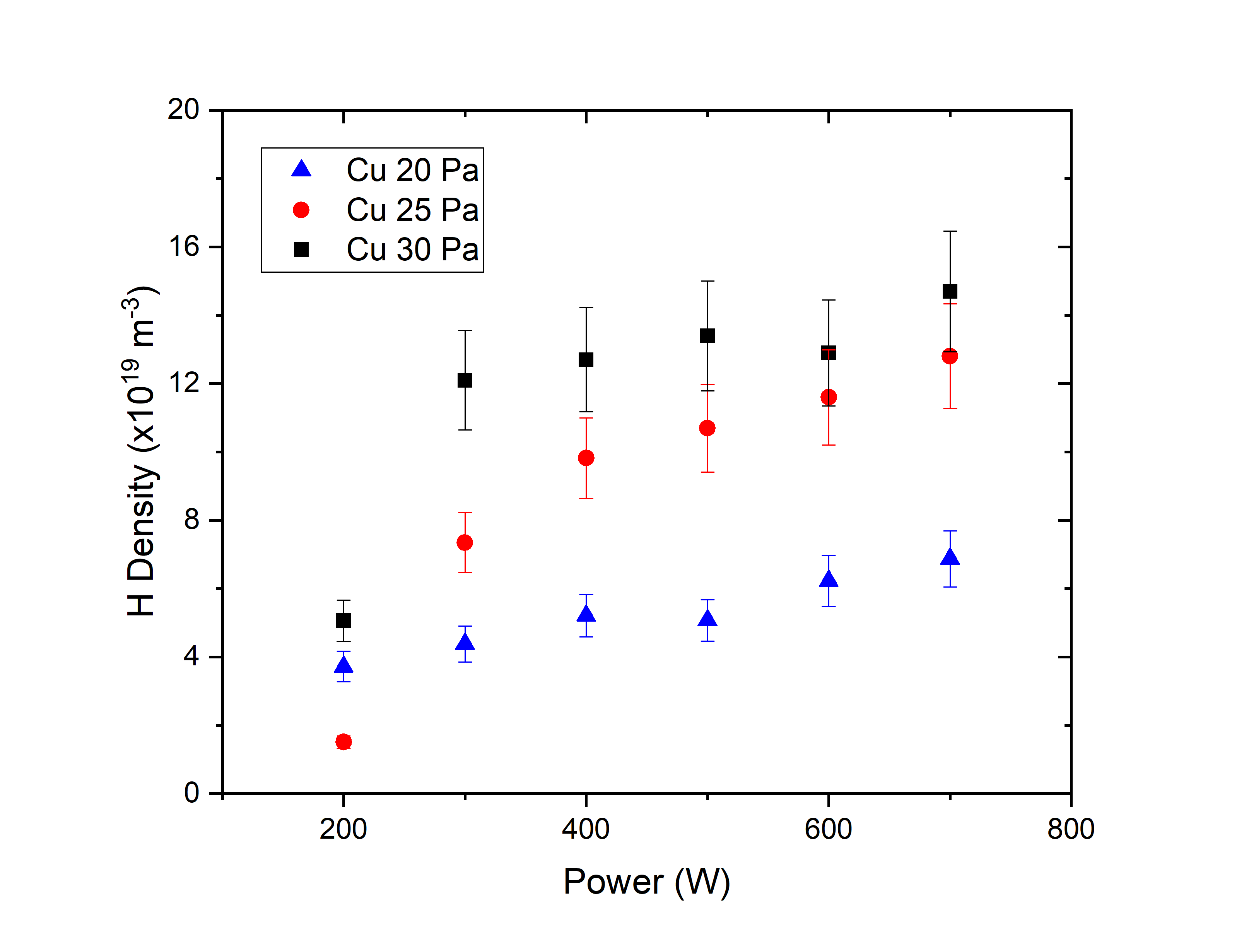}
 \caption{\label{figureeight}Measured H density as a function of applied rf power for the copper surface installed at three different pressures.}
\end{figure}

Figure 7 shows the atomic hydrogen densities, as a function of power for three different pressures, with the stainless steel sample installed. The same pressure dependency can be observed as shown previously in Figure 6. Equally, a similar trend on the atomic hydrogen density with increasing power can be observed in Figure 6 and 7.

\begin{figure}[htb!]
 \centering
 \includegraphics[width=12cm]{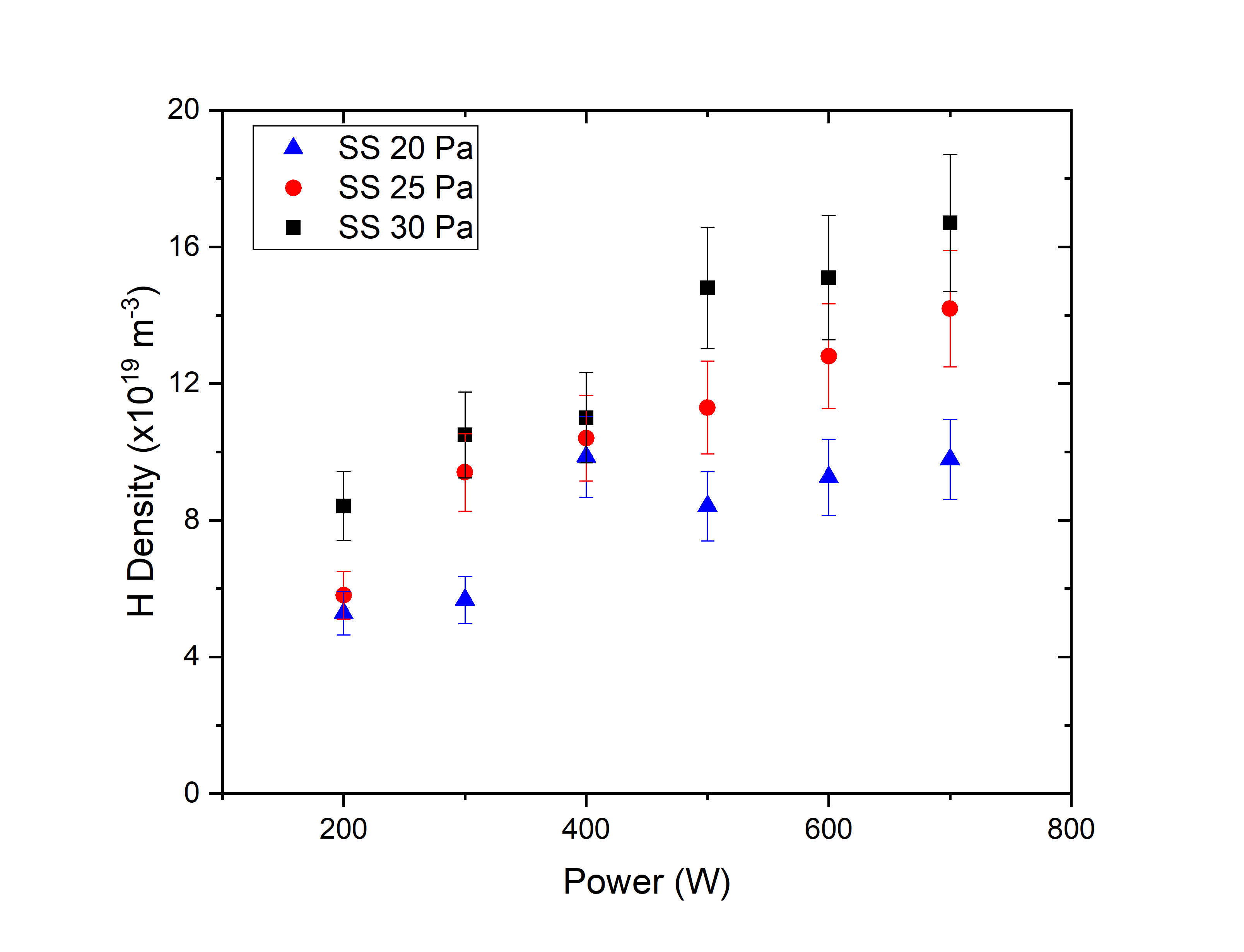}
 \caption{\label{figureeight}Measured H density as a function of applied rf power for the stainless steel surface installed at three different pressures.}
\end{figure}

Figure 8 shows the atomic hydrogen density, as a function of power for the five different materials investigated, at a gas pressure of 25 Pa. The atomic hydrogen density increases as a function of power for all five surfaces, excluding the 500 W measurement for aluminium, with the highest densities being recorded for the copper and steel surfaces. This is in contrast to Figure 3, the H$^{-}$ ion density measurements for the five different surfaces, that showed a clear distinction between the copper and steel.

The increase in the atomic hydrogen density with increasing pressure may explain the decrease in the H$^{-}$ density at higher pressures. However, further investigations are required to ascertain whether this is due to the associative detachment mechanism, or through the relaxation of vibrationally excited hydrogen from atomic collisions.  

\begin{figure}[htb!]
 \centering
 \includegraphics[width=12cm]{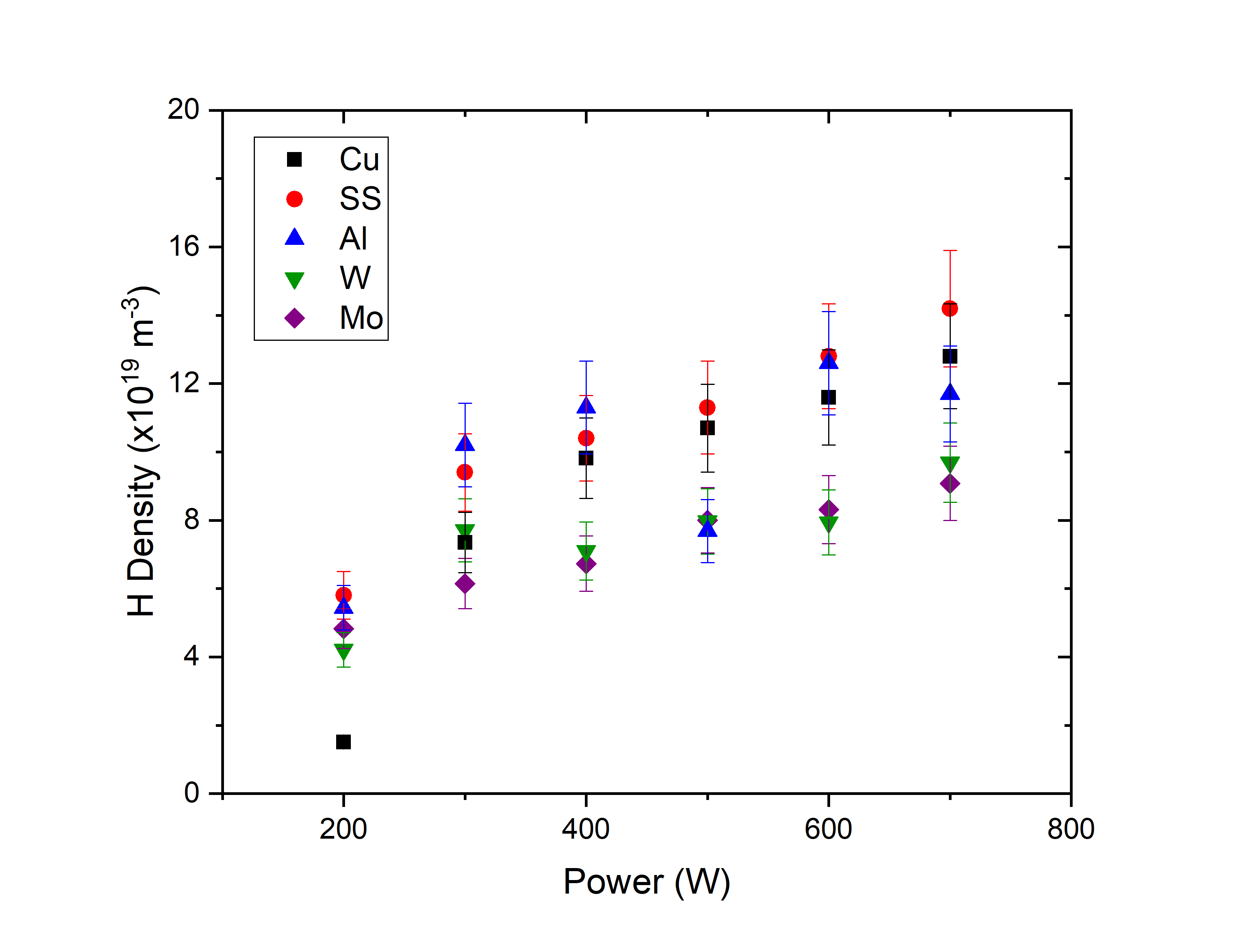}
 \caption{\label{figureeight}Measured H density as a function of applied rf power for different surface materials at a fixed pressure of 25 Pa}
\end{figure}

\newpage

\section{Conclusion}

This paper has studied the production of H$^{-}$ ions in a modified GEC reference cell, using a pulsed radio-frequency inductively coupled plasma source operated in E-mode, with different grounded electrode materials installed. Through the use of laser-photodetachment, we have observed a difference in the volume production of H$^{-}$ ions produced above different metallic surfaces. The most likely cause are the vibrationally excited species produced due to atomic recombination at the surface. These vibrationally excited species can undergo dissociative attachment in order to produce the measured negative ions. Two-photon absorption laser induced fluorescence measurements were conducted to confirm an increased atomic hydrogen density with increasing pressure; this may explain the observed trend of the H$^{-}$ ion density with pressure, due to either associative detachment or V-t relaxation of vibrationally excited hydrogen molecules.   

\section{Acknowledgements}

The authors would like to thank R. Armitage for all their help with the technical support and recognise the funding received through grant EP/K018388/1. J.E would like to thank the EPSRC Centre for Doctoral Training in the Science and Technology of Fusion Energy for affording him this opportunity and funding part of the project through grant EP/L01663X11.

\section{References}

\bibliographystyle{unsrt} 

\bibliography{References} 

\begin{thebibliography}{10}

\bibitem{Shindo_1995}
H.~Shindo, Y.~Sawa, and Y.~Horiike.
\newblock Silicon etching employing negative ion in {{SF$_6$}} plasma.
\newblock {\em Japanese Journal of Applied Physics}, 34(Part 2, No.
  7B):L925--L928, jul 1995.

\bibitem{6990637}
A.~{Aanesland}, D.~{Rafalskyi}, J.~{Bredin}, P.~{Grondein}, N.~{Oudini},
  P.~{Chabert}, D.~{Levko}, L.~{Garrigues}, and G.~{Hagelaar}.
\newblock The pegases gridded ion-ion thruster performance and predictions.
\newblock {\em IEEE Transactions on Plasma Science}, 43(1):321--326, Jan 2015.

\bibitem{KURIYAMA1995445}
M.~Kuriyama, N.~Akino, M.~Araki, N.~Ebisawa, M.~Hanada, T.~Inoue, M.~Kawai,
  M.~Kazawa, J.~Koizumi, T.~Kunieda, M.~Matsuoka, K.~Miyamoto, M.~Mizuno,
  K.~Mogaki, Y.~Ohara, T.~Ohga, Y.~Okumura, H.~Oohara, F.~Satoh, T.~Suzuki,
  S.~Takahashi, T.~Takayasu, H.~Usami, K.~Usui, K.~Watanabe, M.~Yamamoto, and
  T.~Yamazaki.
\newblock High energy negative-ion based neutral beam injection system for
  jt-60u.
\newblock {\em Fusion Engineering and Design}, 26(1):445 -- 453, 1995.
\newblock Proceedings of the Fifth International Toki Conference on Plasma
  Physics and Controlled Nuclear Fusion.

\bibitem{doi:10.1063/1.2814248}
R.~S. Hemsworth, A.~Tanga, and V.~Antoni.
\newblock Status of the iter neutral beam injection system (invited).
\newblock {\em Review of Scientific Instruments}, 79(2):02C109, 2008.

\bibitem{doi:10.1063/1.3541790}
R.~Gutser, C.~Wimmer, and U.~Fantz.
\newblock Work function measurements during plasma exposition at conditions
  relevant in negative ion sources for the iter neutral beam injection.
\newblock {\em Review of Scientific Instruments}, 82(2):023506, 2011.

\bibitem{Babkina_2005}
T~Babkina, T~Gans, and U~Czarnetzki.
\newblock Energy analysis of hyperthermal hydrogen atoms generated through
  surface neutralisation of ions.
\newblock {\em Europhysics Letters ({EPL})}, 72(2):235--241, oct 2005.

\bibitem{Schiesko_2008}
L~Schiesko, M~Carr{\`{e}}re, G~Cartry, and J~M Layet.
\newblock H-production on a graphite surface in a hydrogen plasma.
\newblock {\em Plasma Sources Science and Technology}, 17(3):035023, jul 2008.

\bibitem{doi:10.1063/1.4921298}
M.~Bacal and M.~Wada.
\newblock Negative hydrogen ion production mechanisms.
\newblock {\em Applied Physics Reviews}, 2(2):021305, 2015.

\bibitem{FANTZ20127}
U.~Fantz, P.~Franzen, and D.~W{\"u}nderlich.
\newblock Development of negative hydrogen ion sources for fusion: Experiments
  and modelling.
\newblock {\em Chemical Physics}, 398:7 -- 16, 2012.
\newblock Chemical Physics of Low-Temperature Plasmas.

\bibitem{Franzen_2011}
P.~Franzen, R.~Gutser, U.~Fantz, W.~Kraus, H.~Falter, M.~Fr{\"o}schle,
  B.~Heinemann, P.~McNeely, R.~Nocentini, R.~Riedl, A.~St{\"a}bler, and
  D.~W{\"u}nderlich.
\newblock Performance of multi-aperture grid extraction systems for an
  {ITER}-relevant {RF}-driven negative hydrogen ion source.
\newblock {\em Nuclear Fusion}, 51(7):073035, jun 2011.

\bibitem{FRANZEN20133132}
P.~Franzen, B.~Heinemann, U.~Fantz, D.~W{\"u}nderlich, W.~Kraus,
  M.~Fr{\"o}schle, C.~Martens, R.~Riedl, R.~Nocentini, A.~Masiello, B.~Ruf,
  L.~Schiesko, and C.~Wimmer.
\newblock Commissioning and first results of the {ITER}-relevant negative ion
  beam test facility {ELISE}.
\newblock {\em Fusion Engineering and Design}, 88(12):3132 -- 3140, 2013.

\bibitem{doi:10.1063/1.4932560}
U.~Fantz, B.~Heinemann, D.~W{\"u}nderlich, R.~Riedl, W.~Kraus, R.~Nocentini,
  and F.~Bonomo.
\newblock Towards 20 {A} negative hydrogen ion beams for up to 1 h:
  Achievements of the {ELISE} test facility (invited).
\newblock {\em Review of Scientific Instruments}, 87(2):02B307, 2016.

\bibitem{doi:10.1063/1.5012591}
W.~Kraus, D.~W{\"u}nderlich, U.~Fantz, B.~Heinemann, F.~Bonomo, and R.~Riedl.
\newblock Deuterium results at the negative ion source test facility elise.
\newblock {\em Review of Scientific Instruments}, 89(5):052102, 2018.

\bibitem{Cartry_2017}
G.~Cartry, D.~Kogut, K.~Achkasov, JM. Layet, T.Farley, A.~Gicquel, J.~Achard,
  O.~Brinza, T.~Bieber, H.~Khemliche, P.~Roncin, and A.~Simonin.
\newblock Alternative solutions to caesium in negative-ion sources: a study of
  negative-ion surface production on diamond in {{H$_2$/D$_2$}} plasmas.
\newblock {\em New Journal of Physics}, 19(2):025010, feb 2017.

\bibitem{Sasao_2018}
M.~Sasao, R.~Moussaoui, D.~Kogut, J.~Ellis, G.~Cartry, M.~Wada, K.~Tsumori, and
  H.~Hosono.
\newblock Negative-hydrogen-ion production from a nanoporous {{12CaO}}
  {{7Al$_2$O$_3$}} electride surface.
\newblock {\em Applied Physics Express}, 11(6):066201, may 2018.

\bibitem{doi:10.1063/1.3258352}
L.~Schiesko, M.~Carrère, J.-M. Layet, and G.~Cartry.
\newblock Negative ion surface production through sputtering in hydrogen
  plasma.
\newblock {\em Applied Physics Letters}, 95(19):191502, 2009.

\bibitem{doi:10.1063/1.4841675}
A.~Greb, K.~Niemi, D.~O'Connell, and T.~Gans.
\newblock The influence of surface properties on the plasma dynamics in
  radio-frequency driven oxygen plasmas: Measurements and simulations.
\newblock {\em Applied Physics Letters}, 103(24):244101, 2013.

\bibitem{Greb_2015}
A.~Greb, A.R. Gibson, K.~Niemi, D.~O'Connell, and T.~Gans.
\newblock Influence of surface conditions on plasma dynamics and electron
  heating in a radio-frequency driven capacitively coupled oxygen plasma.
\newblock {\em Plasma Sources Science and Technology}, 24(4):044003, jun 2015.

\bibitem{doi:10.1063/1.4979855}
T.~Tsutsumi, A.~Greb, A.~R. Gibson, M.~Hori, D.~O'Connell, and T.~Gans.
\newblock Investigation of the radially resolved oxygen dissociation degree and
  local mean electron energy in oxygen plasmas in contact with different
  surface materials.
\newblock {\em Journal of Applied Physics}, 121(14):143301, 2017.

\bibitem{Osiac_2007}
M~Osiac, T~Schwarz-Selinger, D~O'Connell, B~Heil, Z~Lj Petrovic, M~M Turner,
  T~Gans, and U~Czarnetzki.
\newblock Plasma boundary sheath in the afterglow of a pulsed inductively
  coupled {RF} plasma.
\newblock {\em Plasma Sources Science and Technology}, 16(2):355--363, mar
  2007.

\bibitem{CHRISTMANN19881}
K.~Christmann.
\newblock Interaction of hydrogen with solid surfaces.
\newblock {\em Surface Science Reports}, 9(1):1 -- 163, 1988.

\bibitem{doi:10.1063/1.467776}
C.T. Rettner.
\newblock Reaction of an {H}-atom beam with {C}l/{A}u(111): Dynamics of
  concurrent {E}ley–{R}ideal and {L}angmuir–{H}inshelwood mechanisms.
\newblock {\em The Journal of Chemical Physics}, 101(2):1529--1546, 1994.

\bibitem{doi:10.1063/1.474302}
S.~Caratzoulas, B.~Jackson, and M.~Persson.
\newblock Eley-{R}ideal and hot-atom reaction dynamics of {H}(g) with {H}
  adsorbed on {C}u(111).
\newblock {\em The Journal of Chemical Physics}, 107(16):6420--6431, 1997.

\bibitem{HARRIS1981L281}
J.~Harris and B.~Kasemo.
\newblock On precursor mechanisms for surface reactions.
\newblock {\em Surface Science Letters}, 105(2):L281 -- L287, 1981.

\bibitem{doi:10.1063/1.480145}
Th. Kammler and J.~K{\"u}ppers.
\newblock Interaction of {H} atoms with {C}u(111) surfaces: Adsorption,
  absorption, and abstraction.
\newblock {\em The Journal of Chemical Physics}, 111(17):8115--8123, 1999.

\bibitem{KAMMLER200091}
Th. Kammler, D.~Kolovos-Vellianitis, and J.~K{\"u}ppers.
\newblock A hot-atom reaction kinetic model for {H} abstraction from solid
  surfaces.
\newblock {\em Surface Science}, 460(1):91 -- 100, 2000.

\bibitem{PhysRevLett.60.337}
R.~I. Hall, I.~\v{C}ade\v{z}, M.~Landau, F.~Pichou, and C.~Schermann.
\newblock Vibrational excitation of hydrogen via recombinative desorption of
  atomic hydrogen gas on a metal surface.
\newblock {\em Phys. Rev. Lett.}, 60:337--340, Jan 1988.

\bibitem{doi:10.1063/1.468242}
C.~Schermann, F.~Pichou, M.~Landau, I.~\v{C}ade\v{z}, and R.~I. Hall.
\newblock Highly excited hydrogen molecules desorbed from a surface:
  Experimental results.
\newblock {\em The Journal of Chemical Physics}, 101(9):8152--8158, 1994.

\bibitem{doi:10.1063/1.3569562}
S.~Markelj and I.~\v{C}ade\v{z}.
\newblock Production of vibrationally excited hydrogen molecules by atom
  recombination on {C}u and {W} materials.
\newblock {\em The Journal of Chemical Physics}, 134(12):124707, 2011.

\bibitem{doi:10.1063/1.1144770}
P.~J. Hargis, K.~E. Greenberg, P.~A. Miller, J.~B. Gerardo, J.~R. Torczynski,
  M.~E. Riley, G.~A. Hebner, J.~R. Roberts, J.~K. Olthoff, J.~R. Whetstone,
  R.~J. Van~Brunt, M.~A. Sobolewski, H.~M. Anderson, M.~P. Splichal, J.~L.
  Mock, P.~Bletzinger, A.~Garscadden, R.~A. Gottscho, G.~Selwyn, M.~Dalvie,
  J.~E. Heidenreich, Jeffery~W. Butterbaugh, M.~L. Brake, M.~L. Passow,
  J.~Pender, A.~Lujan, M.~E. Elta, D.~B. Graves, H.~H. Sawin, M.~J. Kushner,
  J.~T. Verdeyen, R.~Horwath, and T.~R. Turner.
\newblock The gaseous electronics conference radio‐frequency reference cell:
  A defined parallel‐plate radio‐frequency system for experimental and
  theoretical studies of plasma‐processing discharges.
\newblock {\em Review of Scientific Instruments}, 65(1):140--154, 1994.

\bibitem{Abdel_Rahman_2005}
M~Abdel-Rahman, T~Gans, V~Schulz von~der Gathen, and H~F D{\"o}bele.
\newblock Space and time resolved rotational state populations and gas
  temperatures in an inductively coupled hydrogen {RF} discharge.
\newblock {\em Plasma Sources Science and Technology}, 14(1):51--60, jan 2005.

\bibitem{doi:10.1063/1.4931469}
J.~Santoso, R.~Manoharan, S.~O'Byrne, and C.~S. Corr.
\newblock Negative hydrogen ion production in a helicon plasma source.
\newblock {\em Physics of Plasmas}, 22(9):093513, 2015.

\bibitem{PhysRev.28.727}
H.~M. Mott-Smith and Irving Langmuir.
\newblock The theory of collectors in gaseous discharges.
\newblock {\em Phys. Rev.}, 28:727--763, Oct 1926.

\bibitem{Allen_1992}
J.~E. Allen.
\newblock Probe theory - the orbital motion approach.
\newblock {\em Physica Scripta}, 45(5):497--503, may 1992.

\bibitem{Samuell_2015}
M.S. Cameron and C.S. Cormac.
\newblock Low-pressure hydrogen plasmas explored using a global model.
\newblock {\em Plasma Sources Science and Technology}, 25(1):015014, dec 2015.

\bibitem{PhysRevA.67.012707}
T.~Gans, Chun~C. Lin, V.~Schulz-von~der Gathen, and H.~F. D{\"o}bele.
\newblock Phase-resolved emission spectroscopy of a hydrogen rf discharge for
  the determination of quenching coefficients.
\newblock {\em Phys. Rev. A}, 67:012707, Jan 2003.

\bibitem{doi:10.1063/1.476388}
A.~Goehlich, T.~Kawetzki, and H.~F. D{\"o}bele.
\newblock On absolute calibration with xenon of laser diagnostic methods based
  on two-photon absorption.
\newblock {\em The Journal of Chemical Physics}, 108(22):9362--9370, 1998.

\bibitem{doi:10.1063/1.1425777}
M.~G.~H. Boogaarts, S.~Mazouffre, G.~J. Brinkman, H.~W.~P. van~der Heijden,
  P.~Vankan, J.~A.~M. van~der Mullen, D.~C. Schram, and H.~F. Döbele.
\newblock Quantitative two-photon laser-induced fluorescence measurements of
  atomic hydrogen densities, temperatures, and velocities in an expanding
  thermal plasma.
\newblock {\em Review of Scientific Instruments}, 73(1):73--86, 2002.

\bibitem{Niemi_2001}
K.~Niemi, V.~Schulz von~der Gathen, and H~F D{\"o}bele.
\newblock Absolute calibration of atomic density measurements by laser-induced
  fluorescence spectroscopy with two-photon excitation.
\newblock {\em Journal of Physics D: Applied Physics}, 34(15):2330--2335, jul
  2001.

\bibitem{Doyle_2018}
S.J. Doyle, A.R. Gibson, J.~Flatt, T.S. Ho, R.W. Boswell, C.~Charles, P.~Tian,
  M.J. Kushner, and J.~Dedrick.
\newblock Spatio-temporal plasma heating mechanisms in a radio frequency
  electrothermal microthruster.
\newblock {\em Plasma Sources Science and Technology}, 27(8):085011, aug 2018.

\bibitem{Abdel_Rahman_2007}
M~Abdel-Rahman, V~Schulz von~der Gathen, and T~Gans.
\newblock Transition phenomena in a radio-frequency inductively coupled plasma.
\newblock {\em Journal of Physics D: Applied Physics}, 40(6):1678--1683, mar
  2007.

\bibitem{doi:10.1063/1.323539}
H.~B. Michaelson.
\newblock The work function of the elements and its periodicity.
\newblock {\em Journal of Applied Physics}, 48(11):4729--4733, 1977.

\bibitem{doi:10.1063/1.1708797}
R.~G. Wilson.
\newblock Vacuum thermionic work functions of polycrystalline {B}e, {T}i, {C}r,
  {F}e, {N}i, {C}u, {P}t, and {T}ype 304 stainless steel.
\newblock {\em Journal of Applied Physics}, 37(6):2261--2267, 1966.

\bibitem{doi:10.1063/1.1674753}
P.~W. Tamm and L.~D. Schmidt.
\newblock Binding states of hydrogen on tungsten.
\newblock {\em The Journal of Chemical Physics}, 54(11):4775--4787, 1971.

\bibitem{PhysRevA.29.106}
J.~M. Wadehra.
\newblock Dissociative attachment to rovibrationally excited {H}$_{2}$.
\newblock {\em Phys. Rev. A}, 29:106--110, Jan 1984.

\bibitem{doi:10.1063/1.448668}
M.~Pealat, J‐P.~E. Taran, M.~Bacal, and F.~Hillion.
\newblock Rovibrational molecular populations, atoms, and negative ions in {H}2
  and {D}2 magnetic multicusp discharges.
\newblock {\em The Journal of Chemical Physics}, 82(11):4943--4953, 1985.

\bibitem{doi:10.1063/1.1490630}
S.~Gomez, P.~G. Steen, and W.~G. Graham.
\newblock Atomic oxygen surface loss coefficient measurements in a
  capacitive/inductive radio-frequency plasma.
\newblock {\em Applied Physics Letters}, 81(1):19--21, 2002.

\bibitem{Abdel_Rahman_2006}
M~Abdel-Rahman, V~Schulz von~der Gathen, T~Gans, K~Niemi, and H~F D{\"o}bele.
\newblock 15(4):620--626, jul 2006.

\end{thebibliography}

\end{document}